\documentclass[a4paper]{article}
\pdfoutput=1

\usepackage[english]{babel}
\usepackage[utf8x]{inputenc}
\usepackage[T1]{fontenc}
\usepackage{appendix}
\usepackage[group-separator={,}]{siunitx}

\usepackage[a4paper,top=1.96cm,bottom=1.96cm,left=3cm,right=3cm,marginparwidth=1.75cm]{geometry}

\usepackage{amsfonts}
\usepackage{dsfont}
\usepackage{amsmath}
\usepackage{graphicx}
\usepackage{mathtools}
\usepackage[colorinlistoftodos]{todonotes}
\usepackage{hyperref}

\DeclareMathOperator{\sgn}{sign}
\newcommand{\ident}{{\mathds{1}}}

\def\plester{p_{\text{straw}}}
\newcommand\hide[1]{{}}
 \newcommand\anote[1]{\qquad\text{(#1)}}
\title{Biased bootstrap sampling for efficient two-sample testing}

\author{Thomas P. S. Gillam\footnote{GAM Systematic | Cantab}~ and Christopher G. Lester\footnote{University of Cambridge}}
\date{11th November, 2018}
\begin{document}
\maketitle

\hide{Ideas to replace "biased bootstrap"
\begin{itemize}
\item Focused bootstrap
\item Weighted bootstrap
\item Targeted bootstrap
\end{itemize}}

\begin{abstract}
The so-called `energy test' is a frequentist technique used in experimental particle physics to decide whether two samples are drawn from the same distribution. 
   Its usage requires a good understanding of the distribution of the test statistic, $T$, under the null hypothesis. We propose a technique which allows the extreme tails of the $T$-distribution to be determined more efficiently than possible with present methods.  This allows quick evaluation of (for example) $5$--sigma confidence intervals that otherwise would have required prohibitively costly computation times or approximations to have been made.
    Furthermore, we comment on other ways that $T$ computations could be sped up using established results from the statistics community.
    Beyond two-sample testing, the proposed biased bootstrap method may provide benefit anywhere extreme values are currently obtained with bootstrap sampling.
\end{abstract}

\section{Introduction}



The two-sample test statistic, $T$, described in Section~\ref{sec:tsec}, has many uses in particle physics.  The authors of \cite{Barter:2018xbc} seek to address practical problems associated with the calculation of the extreme tails of $T$-distributions.  
This paper develops those themes further, concentrating on two areas:
\begin{itemize}
\item
    We observe that a significant section of the particle physics community dealing with $T$ appears to have become separated from the mathematical statistics literature dealing with Maximum Mean Discrepancy (MMD) metrics, and other integral measures of distributional difference,
even though the two fields are strongly interrelated. 
        \hide{In particular, we
        found in the statistics literature: (i) pre-existing results concerning the scaling relation presented as a conjecture in \cite{Barter:2018xbc} and subsequently investigated by \cite{Zech:2018bpy}, and (ii) a well developed literature concerning so-called `random Fourier features' as calculational devices which can reduce the $O(n^2)$ computational complexity of a naive $T$-statistic computation to $O(n)$.  In Section~\ref{sec:connections} we therefore try to describe these
        connections 
        in
        a manner useful to readers familiar only with the work of \cite{Barter:2018xbc}.}
        In Section~\ref{sec:connections} we describe the connections between these fields, and highlight known results which can directly lead to performance improvements.
\item
    Determining the shapes of the tails of $T$-distributions by simple means is non-trivial and can require prohibitive computational resources. It is to overcome this problem that \cite{Barter:2018xbc} suggests the use of an approximate scaling conjecture so as to obtain approximate knowledge of the $T$-distribution's shape in a sensible time. We note that an alternative way of proceeding is to use a Markov chain to overlay a precisely controlled bias on the 
        vanilla bootstrap procedure (see Section~\ref{sec:bootstrap}) \hide{advocated by \cite{Barter:2018xbc}}
        so that after compensation for this bias, the shape of the tails can become known as precisely as the shape of the core of the distribution.
Such an approach could be useful in any case where those approximations
        are deemed unsuitable or undesirable,
       and could even be used simultaneously for added performance. Our description of this proposal comprises the remainder of this document, from Section~\ref{sec:bootbias} onwards.
\end{itemize}
It is of little practical benefit to be able to generate estimates of the shapes of $T$-distributions if their accuracy cannot be quantified.  Our proposed binned shape estimates are made by un-biasing and re-normalising
histograms that have been filled from Markov chain sequences having internal correlations.  Consequently, the 
uncertainties in every bin
of our shape estimates are non-trivially dependent on each other.  Our procedure for estimating uncertainties in resulting $T$-distributions is therefore itself non-trivial, and is documented in an Appendix
so as not to break the flow of the paper.\footnote{In brief, the Markov chain history is used to generate an estimator for the underlying Markov chain transition matrix, which is itself used to derive the covariance
matrix for every pair of bin counts prior to the weighting and normalisation step. The final uncertainties (after the re-weighting and normalisation process, which introduces further correlations between bins) may then be written as a simple function of the former set of covariances provided that (as has already been assumed) the Markov chain has circulated through most of its domain `a few' times.}  We note, however, that our method of estimating uncertainties 
is considerably faster than the process of calculating the quantities which they themselves constrain.

\hide{It is therefore important to be able to estimate
uncertainties in the shapes of our generated histograms, not only to satisfy readers that we have succeeded in our goals, but so as to provide future users with a means of quantifying the precision of results they might themselves obtain. Since the primary goal is {\it efficient} probing of tails, it is clear that the method of estimating shape uncertainties must itself be no slower than the means of generating distribution's shape, or else the method as a whole is no longer efficient.  To our
horror, our third discovery was that the only
off-the-shelf techniques which appear to be known to the particle physics community for determining histogram shape uncertainties (when that histogram has been filled with correlated samples) are highly computationally inefficient.\footnote{Essentially the only technique used consists of repeating the process of shape determination thousands of times over, and deriving a shape uncertainty from a comparison of the spread of shapes obtained.}  We therefore felt compelled to try to answer the 
question: `Is it possible to determine a closed form expression for the uncertainties in a stochastic histograms filled from Markov chain histories?'.  The answer to this question turns out to be `yes'.  Although we found this independently, we subsequently found evidence in the literature of `Markov chain central limit theorems' that the mathematics underpinning our result has been known for one or more decades. Despite this, so far as we are
aware, that knowledge has not reached end-users (e.g.~particle physicists with histograms generated from Markov chains who are trying to efficiently determine their uncertainties).  The lack of information transfer is perhaps not surprising as words like `statistical uncertainties' and `histograms' are
completely absent from the relevant theoretical works.   Consequently, the final extended goal we set ourselves was to provide a practical guide (see Section~\ref{sec:uncerts}) describing how to efficiently calculate uncertainties in Markov-generated histograms, using knowledge of the history of the events filling them.

In principle, therefore, parts of this paper may be of interest to quite different classes of reader: persons interested in the $T$-statistic (including random Fourier methods and links to maximum mean discrepancy), persons interested in biased bootstrap sampling, and persons interested in uncertainty estimation within histograms of correlated samples.
}

\section{The $T$-statistic}

The `energy test' makes use of a test statistic $T$.
This statistic maps two sets, $S$ and $\bar S$, \label{sec:tsec} to a real number $T(S,\bar S)$, abbreviated $T$. We will refer to each set as a `sample' and refer to the elements of each set as `events'.  The number of events in $S$ is denoted by $n$, and the number in $\bar S$ by $\bar n$.  It is assumed that each event $e_i \in S$ (respectively $\bar e_j \in \bar S$) is independent and identically distributed with underlying unknown distribution $p$
(respectively $\bar p$), i.e. $e_i \sim p$  and $\bar e_j \sim \bar p $.   In common with most two-sample test-statistics, the job of $T$ is to assist in determining whether or not $p$ is the same as $\bar p$.  Summarising crudely: if $p$ is very similar to $\bar p$ then $T$ is expected to take values close to or below zero, while increasing differences between $p$ and $\bar p$ should (at fixed $n$ and $\bar n$) lead to ever greater positive values for $T$. The definition of $T$ used in
\cite{Barter:2018xbc} is given in equation~(\ref{eq:T}),
\begin{align}
T = 
\frac 1 2 \frac 1 {n(n-1)} \sum_{i\ne j}^n \psi(e_i,e_j) 
+
\frac 1 2 \frac 1 {\bar n(\bar n-1)} \sum_{ i\ne  j}^{\bar n} \psi(\bar e_{ i}, \bar e_{ j}) 
-
\frac 1 {n \bar n} \sum_{i, j}^{n,\bar n}\psi(e_i,\bar e_{j})
\label{eq:T}
\end{align}
in which $\psi$ is an appropriately chosen kernel function which maps any two events (whether from $S$, $\bar S$ or both) to a number.   In any real-world use-case, the choice of $\psi$ is probably the single biggest determinant of the usefulness of $T$ as a statistic. Much work should therefore go into choosing $\psi$ appropriately. Our own paper, however, is not interested in the statistical optimality of $T$, but seeks instead to improve the efficiency with which the tails of distribution of $T$-distributions (for any $\psi$) can be sampled.  Consequently, we fix $\psi$ 
to be the same exponential function of a Euclidean distance between events used by \cite{Barter:2018xbc}:
\begin{align}
\psi(e_i,e_j) = \exp\left[- \frac {{|e_i - e_j|}^2} {2 \delta^2}\right],
\end{align}
with $\delta=\frac 1 2$, even though many alternatives choices of $\psi$ are possible.\footnote{Given the decision to fix  $\delta$ at $\frac 1 2$, there is strong motivation to use distributions $p$ and $\bar p$ having length scales of similar or slightly larger order. For this reason our results in Section~\ref{sec:results} use a $p$ which distributes events $e_i$ uniformly within a unit cube.}

Conceptually there is a difference between (a) the single value of $T$ that might be computed from a sample $S$ obtained from a physical detector, together with a reference sample $\bar S$ obtained from somewhere else, and (b) the (hypothetical) $T$-values that could appear under a null hypothesis that $p=\bar p$.  To distinguish between the two
we refer to the former statistic as $T(S,\bar S)$, to indicate that it is a function of the particular samples it digests, and denote the latter random variable by $T(p, n,\bar n)$, to emphasise the values it can take are governed by the model $p$ assumed in the null hypothesis, together with the numbers of events that will populate each sample. Whether talking of the statistic $T(S,\bar S)$ or the random-variable $T(p,n,\bar n)$, equation (\ref{eq:T}) still applies.

When hypothesis testing, it is of central importance that it is possible to evaluate $P(T(p, n, \bar n) \ge \hat T)$, which is the probability that $T(p,n,\bar n)$ exceeds any fixed value $\hat T$. 
If $p$ is independently known, this probability can (in principle) be determined by generating many independent samples $S$ and $\bar S$ from $p$, from which can be recorded the fraction of resulting $T$-values which are sufficiently extreme. 
In practice, there are two reasons such an approach is problematic.  The first problem (`P1') is that the formula for $T$ in equation (\ref{eq:T}) has a complexity which scales as $\max(n^2,\bar n^2)$. This would appear to be bad news for particle-physics, where it is not unusual to need samples with $n$  of order $10^8$.   The second problem (`P2') is that, even if $T$ evaluation is fast, it would be necessary to generate $O(3.5\times 10^{8})$ independent sets $S$ and $\bar S$ and their
associated $T$ values in order to determine the value of $T_0$ corresponding to a 5-sigma excursion with 10\% precision.  If neither issue were addressed, a naive 5-sigma $p$-value calculation in particle physics could thus easily require an infeasible $O(10^{2\times 8+8})$ calculations.

The work of \cite{Barter:2018xbc} could be said to address P1, insofar as it 
observes empirically that, for large enough $n$, there exist values 
of $k$ of order $100$ such that the desired $T(p,n,n)$-distribution is well 
approximated by the distribution of 
$(k / n) T(p,k ,k )$.
Since the complexity of the latter statistic is $O(100^2)$ and 
not $O(n^2)$, it has addressed P1 successfully.

In contrast, our own work addresses P2. That is to say: we seek to find better
ways of sampling the extreme tails of the $T$-distribution (e.g. to calculate 5-sigma or 15-sigma $p$-values), independently of 
whether the $T$-statistic (or an approximation to it) is costly to calculate.  
Our work is thus complementary to that of \cite{Barter:2018xbc}.
Future users may apply both strategies at once, 
or use either separately, depending on their needs.


In passing, we note that
users who need to address P1 (perhaps because they have large $n$), but who cannot use the method of 
\cite{Barter:2018xbc} (perhaps because their $n$ is not large enough to make the scaling-approximation sufficiently accurate) might
consider using
the `random Fourier features' method described in Section~\ref{sec:connections} as it provides an alternate and well-established $O(n)$ technique for calculating values of the $T$-statistic with controlled uncertainty.

\section{Related work}
The above $T$-statistic
was first introduced in 2002-2005 in the High Energy Physics community \cite{Zech2002,Zech2003}, and this method has been cited in several publications from the LHCb collaboration \cite{LHCb2014,LHCb2016}, and in methods pertaining to the same experiment \cite{Williams2011,Parkes2016}.

Separately there have been a series of developments regarding MMD metrics, which also test for discrepancies between two samples. These took off with the work of \cite{Borgwardt2006,Gretton2008,Gretton2012}, but appear to inherit characteristics of a technique originally introduced in 1953 \cite{Fortet1953}. A critical observation is that the energy test given above is equivalent to an MMD test\footnote{Observe that equation~(\ref{eq:T}) above differs from equation~(3) of \cite{Gretton2012} only by a factor of 2.}.

\label{sec:connections}

A formal correspondence between energy distances and MMD is known in the statistics literature \cite{Sejdinovic2012}, however there does not appear to be an appreciation of this link in the High Energy Physics literature. We suspect that recent developments for fast and efficient two-sample tests would be of interest to the LHCb collaboration and others. In particular we would highlight an efficient spectral approximation of the null hypothesis distribution \cite{Gretton2009}, the $B$-test
\cite{Zaremba2013}, and the use of `random Fourier features'  in reducing the computational complexity of MMD test-statistic computation to $O(n)$ from $O(n^2)$ \cite{Zhao2014}.  Our contribution is orthogonal to the techniques mentioned in this section, since we directly target the efficiency of the bootstrap method \cite{Efron1993} itself. We also note that our method can be applied anywhere bootstrap is used, for example in independence testing \cite{Zhang2016}.


\label{sec:barterasspectral}


As has already been mentioned, the authors of \cite{Barter:2018xbc} conjecture that, in the case of equal sample sizes $n=\bar{n}$, the distribution of $nT$ is approximately independent of $n$ for sufficiently large $n$. A proof is not provided therein, rather an appeal to the uncorrelated case is made in which the scaling is known to be exact. Following  \cite{Barter:2018xbc} others \cite{Zech:2018bpy} have investigated the relation in more detail and affirmed the conjecture. We observe, however, that such results have been known for a
decade in other fields.  For example, in 2009 it was shown in \cite{Gretton2009} that there is an asymptotic limit for the distribution of $nT$ as $n\rightarrow\infty$, namely
\begin{align}
nT\xrightarrow[D]{} \frac{1}{2} \sum_{l=1}^\infty \lambda_l (z_l^2 - 2),\nonumber
\end{align}
where $z_l\sim \mathcal{N}(0, 2)\ \forall l$, and $\lambda_l$ represents the $l^\textrm{th}$ eigenvalue in the following equation
\begin{align}
\int_\mathcal{E}\tilde{\psi}(e_i, e_j) \phi_l(e_i) dp = \lambda_l \phi_l(e_j).\nonumber
\end{align}
In this context, $\phi_l$ are the eigenfunctions, and $\tilde{\psi}(e_i, e_j) = \psi(e_i, e_j) - \mathbb{E}_e\psi(e_i, e) - \mathbb{E}_{e,e'}\psi(e, e')$.\footnote{In practice these eigenvalues can be estimated empirically for finite sample size, and Section~2 of \cite{Gretton2009} details this method.}  Not only does this confirm the conjecture of \cite{Barter:2018xbc}, as does \cite{Zech:2018bpy}, it goes further by providing an expression for the limiting distribution for
arbitrary kernel function.

\section{Bootstrap sampling}
\label{sec:bootstrap}

Suppose one has a desire to calculate properties of the $T(p,n,\bar n)$-distribution when in possession of incomplete or limited information about $p$. Specifically, suppose that knowledge of $p$ is limited to the availability of one sample $S_{m}$ containing $m$ events, each independently sampled from $p$\footnote{%
Typically one might have $m=n+\bar{n}$, in the case in which one considers the \emph{union} $S\cup\bar{S}$ as the set of events from which to draw bootstrap samples.
}.  In such a situation, bootstrap sampling can help. Bootstrap sampling \cite{bootstrap1} makes use of an approximation ${\tilde p}$ to the unknown distribution $p$ made by combining $m$
delta distributions, one centred on each of the events in $S_{m}$:
$$\tilde p(e ) = \frac 1 {m} \sum_{e_i \in S_{m}} \delta{(e-e_i)}.$$ This distribution $\tilde p$, sometimes referred to as the `empirical distribution for $p$', places probability only at the locations where events have already been seen.  With $\tilde p$ so defined, samples drawn from $T(\tilde p, n,\bar n)$ then become nothing other than values of T generated from $n$ and $\bar n$-sized collections of events drawn independently from $S_{m}$ with replacement.  Any of the desired properties of $T(p,n,\bar n)$ may then be estimated by repeated
sampling from $T(\tilde p,n,\bar n)$.

While bootstrap sampling provides nothing more than an {\it estimate} of any of the desired properties of $T(p,n,\bar n)$, the literature \cite{doi:10.1093/biomet/68.3.589} suggests that differences between estimates and underlying parameters become negligible for our purposes when the number of event $\min\left[n,\bar n, m\right]$ exceeds 50 or 100.  
For the purposes of this paper we therefore make the assumption (in keeping with \cite{Barter:2018xbc}) that bootstrap sampling (rather than `true' sampling) introduces only negligible uncertainties in all the places we use it.  While it is legitimate to question the validity of this assumption, doing so is not relevant for the purposes of comparing our approach to that of \cite{Barter:2018xbc}.

\section{Efficient sampling in the $T$-tails}
\label{sec:bootbias}

Unavoidably, generators of uncorrelated samples produce extreme samples only rarely.  It has already been mentioned that this is what makes it computationally inefficient to generate multi-sigma tails of distributions using vanilla bootstrap sampling.  We propose that, to overcome this difficulty, the independent bootstrap sampler of \cite{Barter:2018xbc} be replaced with a Metropolis-Hastings Markov chain that: (a) generates {\bf correlated} samples via a random walk on a state-space of
bootstrap samples, and (b) has a calculable {\bf bias} toward extreme values.  The bias helps the chain to find the tails, while the correlations permit the tails (once discovered) to be explored more thoroughly.

{\bf Correlation} between bootstrap samples in the Markov chain may be introduced in various ways. The simplest method is to require that the  `proposal distribution', $\hat Q$, 
which suggests Markov-state transitions from state $s_i$ to state $s_{i+1}$, should update only a fixed fraction of the events forming each bootstrap sample.  For our purposes, we found it sufficient to use a $\hat Q$ which leaves a random 90\% of the events in each bootstrap samples unchanged, and draws the
remaining 10\% of events as fresh samples from the relevant empirical distribution.  Users working on other problems may find it necessary to investigate alternative choices.\footnote{While the Metropolis-Hastings method is well defined for a wide class of proposal functions, poor choice of proposal function could easily lead to very inefficient or incomplete sampling.  We found our sampler to be stable with respect to large changes in this 90:10 ratio, but others using this method will  have to determine what sort of proposal function has,
for their application, the right balance between high variability (to allow state-space exploration) and correlation (to allow benefit to be obtained from information learned). By allowing proposals to change around 10\% of the events in any one Markov `step' the sampler can, in principle, reach an entirely independent bootstrap sample in order ten successful steps.  By this measure, our proposal function has a relatively short memory.}  This particular proposal algorithm $\hat Q$  happens to have a density, $Q$,  which is symmetric: $Q(s_{i+1}\vert s_i)=Q(s_i\vert s_{i+1})$.

{\bf Bias} toward extreme values of $T$ may be obtained by modifying, with a weighting-function $f(T)$, the ratio $\rho$ of $Q$-factors used in the Metropolis-Hastings acceptance rule.  Specifically: a proposal to move to a new Markov state $s_{i+1}$ given a current state $s_{i}$ is made.  The proposal is accepted only if a random number chosen uniformly from the real interval between 0 and 1 is found to be less than the ratio $\rho$, where:
$$
\rho = \frac{Q(s_{i}\vert s_{i+1})}{Q(s_{i+1}\vert s_i)} \frac{f(T(s_i))}{f(T(s_{i+1}))}.
$$
If not accepted, the existing state is instead re-asserted (i.e.~$s_{i+1}$ is set equal to $s_i$). Since our $Q$ is symmetric, its density cancels in the definition of $\rho$ above and so does not  need to be evaluated.
If $p(T)dT$ is (by definition) the desired $T$ distribution, then with the above rule the random walk intrinsically samples from the weighted density $p(T)f(T)dT$, and so the resulting $T$-values generated by the chain must be unweighted\footnote{By `unweighted' we mean `weighted with weight $1/f(T)$'.} to bring them back to the desired distribution.  The considerable freedom in choice of $f(T)$ is highly constrained by the improvements desired. If the goal of the biasing is to obtain quantitative information about the tails, a minimal requirement would be that $f(T)$ allow {\bf efficient} recirculation of the Markov Chain between the bulk and the tail regions.  Without this, co-normalisation of the tail with respect to the bulk will not be achieved.  Practically speaking this means that $f(T)$ must be approximately proportional to $1/p(T)$ so that the Markov chain will visit all $T$-values approximately uniformly.  A measure of the degree of departure of $f(T)$ from $1/p(T)$ could be taken to be the largest factor, $K$, by which one part of the approximately uniform density exceeds another, within a subset $\tau$ of the interesting range of $T$-values:
\begin{align}
K(\tau)
=
\exp\left(\max_{T_1,T_2\in \tau}\left|
\log\left(
\frac
 {f(T_1)p(T_1)} 
 {f(T_2)p(T_2)} 
\right)
\right|\right)\ge 1.\label{eq:kisdefined}
\end{align}
As a rule of thumb: keeping $K$ below $\sim 10$ in interesting regions of $T$ could be considered a first-order requirement for efficiency. Values of $K$ larger than this would indicate that there are parts of the $T$-distribution in which the Markov chain is lingering for an order of magnitude more time than necessary, meaning that a better choice of $f(T)$ might have allowed equally good precision for an order of magnitude less computation time.\footnote{Note that it is neither possible nor desirable to find a weight function that is {\it exactly} proportional to $1/p(T)$ and so having $K=1$.  To begin with, the existence of such a weight would remove the
need to use the sampler since its only purpose is to estimate $p(T)$.   Furthermore, it is not possible to sample from an unbounded uniform distribution -- so some departure from $f(T)\approx 1/p(T)$ will always be needed outside the range of interesting $T$ values $\tau$. In such regions the simplest prescription is to replace $f(T)$ by a constant. This causes the sampler to return to unbiased behaviour.}

Since the efficiency of the sampler is strongly dependent on the choice of weight-function $f(T)$ it must be choosen carefully.  
Our approach is to identify a class of functions whose tails have similar properties to the tails of $T$-distributions, and which are completely specified by their mean, second central moment and third central moments.\footnote{Note that is is not necessary to use functions be determined by their moments. Any parametric or non-parametric function that can
approximately fit a $T$-distribution and can be extrapolated non-pathologically toward the tails could be used.  We base our approach on moment fitting simply as it provides for a non-iterative fitting with constant-time guarantees.} 
We then use a very short pre-sampling phase (that makes no use of weight-functions for tail boosting) to gain crude estimates of the mean, second central moment and third central moment of our desired $T$-distribution. The member of the aforementioned class of functions
having the same moments is then identified,
and its reciprocal (at least in the $T$ range of interest) is used as the final weight function $f(T)$.
Inevitably the moment estimates obtained from such a pre-run are unlikely to be optimal: the pre-run is unlikely to sample far into the tails, and third moments are notoriously unstable things to estimate from finite samples. Nonetheless, the bar is low for tolerability; a weight function $f(T)$ will speed things up if it is not pathological and is a better approximation to the desired function (in our case $1/p(T)$) than is a constant.  Both are relatively easy to achieve.\footnote{Though we have not found the need to
do so, one could iterate this whole process if the resulting weight functions were found to be of insufficient quality.  For example, a short pre-run without weighting could be used to find a first guess at an optimal weight function, and this could be followed by a secondary pre-run using that first guess to find a more accurate second guess, etc.}


\section{Straw models for $T$-tails}

\begin{figure}
\centering
\includegraphics[width=0.6\textwidth]{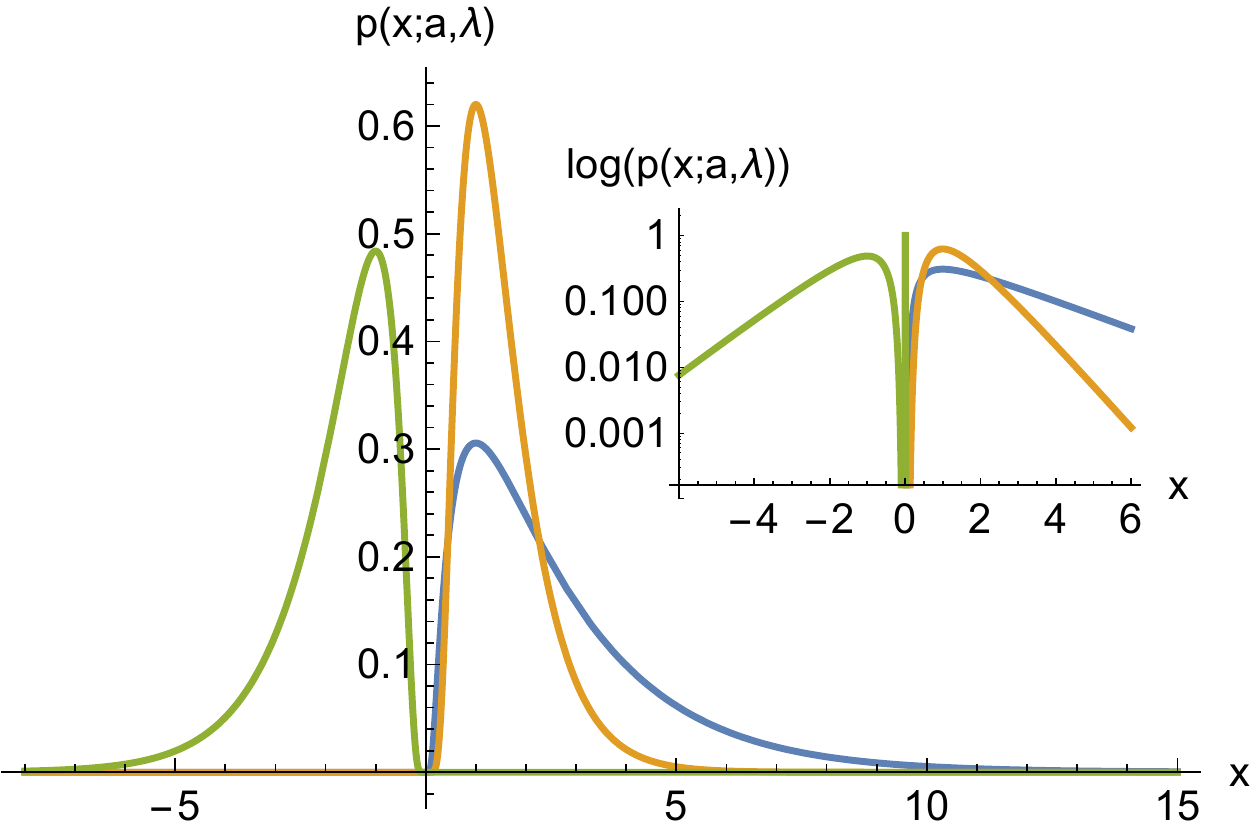}
    \caption{\label{fig:lesterdistribs} Examples of the distributions $\plester(x;a,\lambda)$ defined in Equation~(\ref{eq:lesterdistrodef}). Values of $(a,\lambda)$ shown are:  $(1,  1)$ in blue, $(1, 3)$ in dirty yellow, and $(-1, 2)$ in green.}
\end{figure}

Strictly speaking, $T$-distributions are discrete and with bounded support. At large $n$, however, for all practical purposes they are well approximated by continuous distributions with support on a semi-infinite region of the real line. We make the heuristic observation that the logarithm of the probability density function is often well approximated by a hyperbola with a vertical asymptote on one side and an oblique asymptote on the other. Examples of such hyperbole are shown in the
inset plot within Figure~\ref{fig:lesterdistribs}. This motivates considering a class of probability densities with similar properties.  The simplest such class will have three parameters: one degree of freedom being needed to control the location of the
vertical asymptote, while the other two specify the location and slope of the oblique asymptote.  It is not necessary that the three parameters have exactly those roles, only that there be three degrees of freedom capable of providing the necessary control.  Indeed, since an overall location parameter may be trivially introduced into any distribution that lacks one,
it is sufficient to look for two-parameter distributions
whose location is not controlled.

\def\mainFigACurveOneNumEvents{200}
\def\mainFigACurveOneNumSamples{25000}
\def\mainFigACurveOneNumPreSamples{1000}
\def\mainFigACurveOneNumBins{47}
\def\mainFigACurveOneBinWidth{0.0004872340425531915}
\def\mainFigACurveOneTMin{-0.0027}
\def\mainFigACurveOneTMax{0.0202}
\def\mainFigACurveOneDistroName{distro\_uniform\_in\_cube}
\def\mainFigACurveOneSigmaLevel{(1,)}
\def\mainFigACurveOneDeltaInPhi{0.5}
\def\mainFigACurveTwoNumEvents{200}
\def\mainFigACurveTwoNumSamples{25000}
\def\mainFigACurveTwoNumPreSamples{None}
\def\mainFigACurveTwoNumBins{47}
\def\mainFigACurveTwoBinWidth{0.0004936170212765957}
\def\mainFigACurveTwoTMin{-0.003}
\def\mainFigACurveTwoTMax{0.0202}
\def\mainFigACurveTwoDistroName{distro\_uniform\_in\_cube}
\def\mainFigACurveTwoSigmaLevel{(1,)}
\def\mainFigACurveTwoDeltaInPhi{0.5}
\def\mainFigACurveThrNumEvents{200}
\def\mainFigACurveThrNumSamples{10000000}
\def\mainFigACurveThrNumPreSamples{None}
\def\mainFigACurveThrNumBins{47}
\def\mainFigACurveThrBinWidth{0.0004936170212765957}
\def\mainFigACurveThrTMin{-0.003}
\def\mainFigACurveThrTMax{0.0202}
\def\mainFigACurveThrDistroName{distro\_uniform\_in\_cube}
\def\mainFigACurveThrSigmaLevel{(1,)}
\def\mainFigACurveThrDeltaInPhi{0.5}

\begin{figure}
\centering
\includegraphics[width=0.9\textwidth]{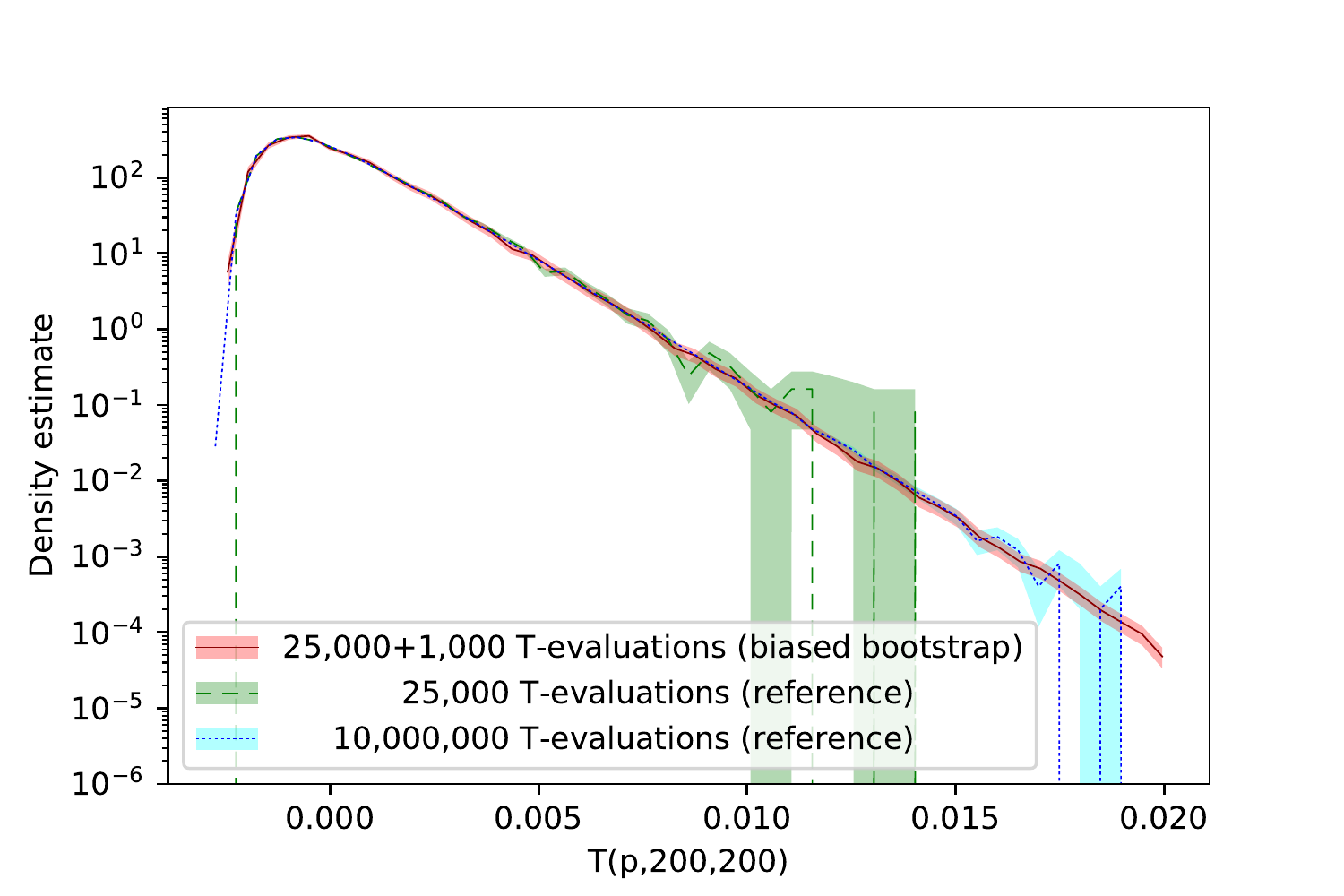}
    \caption{\label{fig:mainfiga} 
    We show a number of different estimates of the $T(p,n,n)$ density
    for $n=\num{\mainFigACurveOneNumEvents}$ events
    assuming a distribution $p$  
    which distributes events uniformly within a unit three-dimensional cube.
    One estimate (shown in solid red) is generated from $N=\num{\mainFigACurveOneNumSamples}$ of our own biased bootstrap MCMC samples and $\num{\mainFigACurveOneNumPreSamples}$ unbiased initialisation samples. It may be compared with the estimate obtained from $N=\num{\mainFigACurveTwoNumSamples}$ unbiased bootstrap non-MCMC samples (dashed green), and $N=\num{\mainFigACurveThrNumSamples}$ unbiased bootstrap non-MCMC samples (dotted blue). 
    All estimates use $\mainFigACurveOneNumBins$ bins. In each bin, $b$, the density is shown as the central line, while the shaded region shows the one-sigma uncertainty on the density, as calculated by the method documented in the Appendix.
    Technically, both the densities and uncertainties are discrete and represent averages over each bin. However, for the purposes of display only they have been made to look continuous (i.e.~the values have been placed at the centre of each bin and joined to neighbours with straight-line
    interpolation) as this makes the overlapping regions easier to follow.
    }
\end{figure}

Being initially unable to find an off-the-shelf distribution with the above properties, we proposed one of our own.\footnote{We initially considered the Generalised Extreme Value (GEV) distribution suggested by \cite{Barter:2018xbc} as a candidate model for the tails, but dismissed it after noting that  its tails do not, in general, fall as the exponential of a linear function of $T$, and so these usually lead to large and inefficient values of $K$ (as defined in equation (\ref{eq:kisdefined})) in our test cases. After settling on our straw model, however, we have noted that the a special cases of the GEV called the Gumbel distribution has a probability density function with a tail of the right form. It is potentially possible, therefore, that the Gumbel could have been used in place of our own straw model. We have not investigated this possibility further, but note that as the Gumbel has one fewer parameter than our model, it is by no means guaranteed that it has sufficient flexibility to fit to both the left and right hand sides of the $T$-distribution simultaneously.}  The simplest we could imagine has the following unit-normalised probability density function:
\begin{align}
\plester(x; a,\lambda)=\begin{cases}
\frac{
  \exp\left[{
    -\frac \lambda 2 \left(
      \frac x a + \frac a x 
\right)}\right]
}
{2 |a| K_1(\lambda)}
& \text{if $a x > 0$}\\
0 & \text{otherwise,}
\end{cases}\label{eq:lesterdistrodef}
\end{align}
in which $K_n(z)$ is a modified Bessel function of the second kind. 
It is controlled by a scale parameter $a$ (which is also the mode of the distribution) and a skewness parameter $\lambda$.
The parameter $\lambda$ is always required to be greater than zero, while $a$ must be non-zero. The distribution is one-sided, with support on $(0,\infty)$ if $a$ is positive, and on $(-\infty,0)$ if $a$ is negative. 
Some examples of this distribution are shown in Figure~\ref{fig:lesterdistribs}, in which the inset part of the figure demonstrates the desired hyperbolic properties of the tails.

After the trivial addition of a translational degree of freedom, it is the above density that becomes our idealised $T$-distribution's `model', and whose reciprocal (in the $T$-regions in which we are interested) that becomes our weight function $f(T)$. 
The process of fitting the above model to a set of $T$-values from a pre-run could, in principle, be done by many different methods.  As we are desirous of obtaining a reliable fit in constant time, we eschew iterative solution finders and instead compute three moments of the data set that is being fitted. In the first part of the Appendix we have supplied  expressions for the quantities $\lambda$ and $a$ (and the missing position parameter) in terms of those first three moments.  These expressions,
together with the moments already calculated, are what we use to perform our fit to pre-run data.  We emphasise, however, that many other approaches would be possible that rely on none of those results.

Finally, we note that it could be necessary to invent or use other straw models in any cases where $T$-distributions were found to look wholly unlike the forms shown in Figure~\ref{fig:lesterdistribs}. Any mis-match between straw model and real distribution ought to show up readily through poor efficiency and increasing uncertainties on the resulting estimates, and so this area should be largely self-policing.

\section{Resulting $T$-distributions}
\label{sec:results}

Figure~\ref{fig:mainfiga} compares three different estimates of the shape of the $T(p,200,200)$ distribution for a $p$ which distributes events uniformly within a three dimensional unit cube. Two of the shape estimates (green-dashed and blue-dotted) are generated by a vanilla bootstrap process and act as `reference' distributions against which the estimate from our own `biased bootstrap' estimate (red-solid) can be compared.  The `biased bootstrap' estimate is generated from \num{\mainFigACurveOneNumSamples}\ evaluations of $T$,
following \num{\mainFigACurveOneNumPreSamples}\ pre-samples (used to evaluate parameters of the bias function).  This is the same number of $T$-evaluations used by the green-dashed reference distribution, and so a comparison between the two (between red-solid and green-dashed) provides a fair assessment of the performance benefits of `biased bootstrap' over the reference method.  It can be seen that the red and green estimates are in agreement with each other over their common
domain, but that the uncertainty of the green reference method increases dramatically once the density falls to $10^{-3}$ of its height at the peak.  The green estimate is evidently useless  above $T$-values of order 0.010, while the `biased bootstrap' estimate appears well controlled up to $T=0.020$ (and presumably beyond).  To check that the biased-bootstrap prediction is correct beyond the point at which the statistics of the green expire, one can compare it to the dotted blue line, which is
generated with three orders of magnitude more statistics (\num{\mainFigACurveThrNumSamples} $T$-evaluations).  Not only is the agreement between red and blue confirmed to be good everywhere, the benefits of the `biased bootstrap' approach are once again made manifest. The biased method continues to work well above $T$-values of 0.016, and over seven orders of magnitude of variation in probability density, while the reference method (even with the advantage of 1000 times more
computing power) is unable to match this.

\section{Conclusion}

A biased bootstrap procedure for determining the shapes of $T$-distributions has been described. It has been shown to require many orders of magnitude fewer $T$-evaluations than existing methods when probing distribution tails.  The proposed method does not make approximations beyond the use of bootstrap itself. When desired, it may be used alongside existing methods, such as those in \cite{Barter:2018xbc}, for added speed gains.

\section*{Acknowledgements}
We gratefully acknowledge feedback from William Barter, Rupert Tombs and an anonymous referee acting for the Journal of Instrumentation.  CGL acknowledges support from the United Kingdom's Science and Technology Facilities Council (STFC) consolidated grants RG79174 (ST/N000234/1) and RG95164 (ST/S000712/1).
\section*{Appendix}

\begin{appendices}
\subsection*{Moments and other properties of the $T$-tail straw model}

\begin{figure}
\centering
\includegraphics[width=0.5\textwidth]{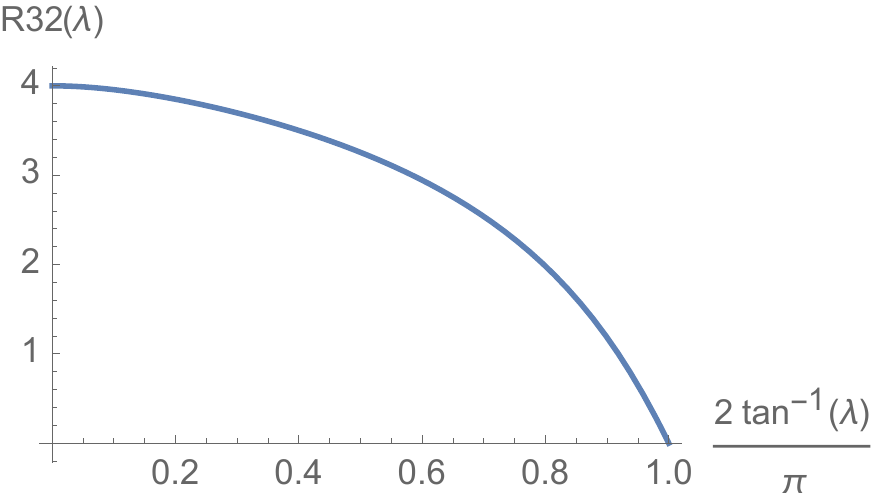}.
\caption{\label{fig:lambdaplot} The dependence of $R_{3,2}(\lambda)$ on $\lambda$.}
\end{figure}

The analytic properties of $\plester(x; a,\lambda)$, which we make use of when fitting to samples obtained in the `pre-run' previously mentioned are as follows:
\begin{itemize}
\item
    The mean takes the value $\displaystyle 
        \left< x\right>=\frac {a K_2(\lambda)}{K_1(\lambda)}$.\vspace{-2mm}
\item
    The $n$\textsuperscript{th} non-central moment of this distribution is:
$\displaystyle\left<x^n\right>= \frac{a^n K_{n+1}(\lambda)}{K_1(\lambda)}.$
\item The 2\textsuperscript{nd} central moment of this distribution is:
\begin{align}
    M_2 &\equiv\left<(x-\left< x \right>)^2\right> \hphantom{This text aligns M_2 with M_3 a few lines below.}\nonumber\\
    &= \left< x^2\right>-2 \left< x\right>\left< x\right>+\left< x\right>^2 \nonumber \\
&= \left< x^2\right>-\left< x\right>^2 \nonumber \\
&= 
\frac{a^2 K_3(\lambda)} {K_1(\lambda)} 
-
\left(\frac{a K_{2}(\lambda)} {K_1(\lambda)}\right)^2 
\nonumber \\
&= \frac{a^2}{K^2_1(\lambda)}
\left(
K_3(\lambda) {K_1(\lambda)} 
-
K^2_{2}(\lambda) 
\right).
\nonumber
\end{align}
\item
    The 3\textsuperscript{rd} central moment of this distribution is:
\begin{align}M_3 &\equiv\left<(x-\left< x\right>)^3\right> \nonumber \\
&= \left< x^3\right>-3 \left< x^2 \right> \left< x\right> + 3 \left< x \right> \left< x\right>^2 - \left< x\right> ^3\nonumber \\
&= \left< x^3\right>-3 \left< x^2 \right> \left< x\right> + 2  \left< x\right>^3 \nonumber \\
&= 
\frac{a^3 K_4(\lambda)} {K_1(\lambda)} 
-
3 \left(\frac{a^2 K_{3}(\lambda)} {K_1(\lambda)}\right) \left( \frac{a K_2(\lambda)}{K_1(\lambda)} \right) 
+
2  \left( \frac{a K_2(\lambda)}{K_1(\lambda)} \right) ^3
\nonumber \\
&= \frac{a^3}{K^3_1(\lambda)}
\left( 
K_4(\lambda) {K_1^2(\lambda)}
-
3 K_3(\lambda)K_2(\lambda)K_1(\lambda)
+
2K^3_2(\lambda)
\right).
\nonumber
\end{align}
\item
The ratio $R_{3,2}(\lambda)$, defined as follows:
$$
R_{3,2}(\lambda) \equiv 
\frac{M_3^2}{M_2^3}
=\frac{
K_4(\lambda) {K_1^2(\lambda)}
-
3 K_3(\lambda)K_2(\lambda)K_1(\lambda)
+
2K^3_2(\lambda)
}
{
K_3(\lambda) {K_1(\lambda)} 
-
K^2_{2}(\lambda) 
}
$$
is a monotonic decreasing function of $\lambda$, whose image is the interval $(0,4)$ as shown in Figure~\ref{fig:lambdaplot}.
\end{itemize}

Using the above results, the moment fitting description can now be described.



\begin{enumerate}
\item
    From the set of $T$-values obtained in the `pre-run', compute the sample mean ($\tilde\mu$), an unbiased estimator ($\tilde M_2$) for the second central moment $M_2$, and an unbiased estimator ($\tilde M_3$) for the  third central moment $M_3$ of the underlying distribution.  Define the constant $\rho={\tilde M^2_3}/{\tilde M_2^3}$.
\item
Numerically, find the value of $\lambda$ which solves the equation
$
R_{2,3}(\lambda)=\rho
$
and call it $\hat \lambda$.
        If and only if the ratio $\rho$ is between 0 and 4, the monotonic and bounded nature of $R_{2,3}(\lambda)$ guarantees a unique solution that is easy to find by bisection search.  If $\rho$ is equal to zero or greater than four, then the estimators are not yet accurate enough, or the distribution being sampled is not well modelled by our straw model.\footnote{If the former, then more statistics in the pre-run are needed. If the latter, then a
        better straw model is needed.}
\item
Define the constant $\hat a$ by the two requirements:
$$|\hat a|= \sqrt{\frac{\tilde M_2 K_1^2(\hat \lambda)}{
K_3(\hat \lambda)K_1(\hat \lambda)-K_2^2(\hat \lambda)
}}$$
and
$$\sgn \hat a = \sgn \hat M_3.$$
Note that the radicand is guaranteed to be positive for any $0< \hat \lambda<\infty$ and $0<\hat M_3<\infty$.
\item
    The `fitted' straw-model for the pre-run's $T$-distribution may finally be defined to be:
        $$p_{\rm Fit}(T) \equiv \plester\left(T-\tilde   \mu+ \frac{\hat a K_2(\hat \lambda)}{K_1(\hat \lambda)};\hat a, \hat \lambda\right)$$ since this distribution has the desired tail shapes, while sharing  mean, second and third central moments with the estimates obtained from the pre-run. The reciprocal of $p_{\rm Fit}(T)$ is what will (in the relevant $T$-range) be used as the weight function to bias the Markov chain.
\end{enumerate}

\subsection*{Uncertainty estimates}
Our density estimates are made by un-biasing and normalising histograms that have been filled from Markov chain sequences having internal correlations.\footnote{The exact formula for our weighting and normalising procedure us given later in (\ref{relntodiff}).} It is very helpful to be able to determine the uncertainties on the resulting density predictions in each bin quickly, from whatever Markov chain run generates the central values of the density estimate.  More formally:
\begin{itemize}
    \item
        Suppose that there exists a mechanism, $M$, that can generate a length-$N$ sequence of values, $T_1, T_2, \ldots, T_N$, as outputs of a Markov chain with fixed (but unknown) transition matrix $P$.\footnote{It is relevant that $P$ is unknown because, although the biased-bootstrap process is a Markov chain, it is one implemted using a Metropolis-Hastings algorithm and therefore has no direct knowledge of its underlying transition matrix, even though one exists in principle.}
    \item
        Suppose that each of the random variables $T_i$ takes a value in a discrete set of indices which, without loss of generality, will be taken to be integers between 1 and $B$ inclusive, and which will be referred to as `bins' for short.\footnote{Since the underlying $T$-values are not binned, some approximation is introduced here.  We have not been able to derive  a generally applicable estimate of the size of the effect introduced by this approximation, though we
        believe it to be small in the cases that matter to us.  There is is room here for 
       further work.}
    \item
        Suppose that $\vec t$ is a particular instance of such a sequence:  $\vec t=\{t_1,t_2,\ldots,t_N\}$, sampled by the mechanism $M$ from the Markov chain governed by $P$.
    \item
        Suppose that the symbol `$s_b$' is used to denote the number of occurrences of the value $b$ in the sequence $t_1, t_2, \ldots t_N$, so that $\vec s = \{s_1,s_2,\ldots s_B\}$ may represent counts in the bins of a histogram.
    \item
        Suppose that $\vec s'' = \{s''_1,s''_2,\ldots s''_B\}$ represents a potentially non-linear transformation of the bin counts in $\vec s$. Equivalently, suppose that there are $B$ functions $g_1, g_2, \ldots, g_B$ such that  $s''_b=g_b(s_1, s_2,\ldots s_B)$. [This non-linear transformation will represent the re-weighting and normalisation steps required in our density estimation.]
    \item
        Suppose that the quantities $\vec S$, $\vec S''$, $S_b$ and $S''_b$ are defined the same way as $\vec s$, $\vec s''$, $s_b$ and $s''_b$ above,  but represent random variables over the process governed by $P$, rather than a particular sample generated therefrom by the mechanism $M$.
\end{itemize}
Given those assumptions, we may pose two questions whose answers assist in the determining the uncertainties of our density estimates.  These questions are:
\begin{enumerate}
    \item
        What are the values of $
\mathrm{Var}[S''_1],
\mathrm{Var}[S''_2],\ldots,
\mathrm{Var}[S''_B]$ in terms of $P$? 
    \item
        If $M$ exists but $P$ is not known, is it possible to write down an estimator for $P$ based on only the single sample $\vec t$, and is this estimator `good enough' to use in place of $P$ in calculating the above variances?
\end{enumerate}
The expression `good enough' is important and relevant even when answering the first question because it is meaningless to talk about `exact' uncertainties.  An uncertainty exists to provide an approximate measure of how  close a prediction might be to an ideal underlying value.  Asking how uncertain the uncertainty is, would lead to a never-ending descent into the uncertainties on uncertainties on uncertainties, {\it etc}.  What is needed, therefore, is a pragmatic approach to
answering the above questions that ensures the answer is fit for purpose, and no more.

Using this freedom, we choose to restrict our attention to cases in which the Markov chain has begun to reach equilibrium.\footnote{In loose terminology we assume the chain has made at least `a few' effectively independent visits to each bin of the space. Equivalently we may require that the chain has lost knowledge of its starting point `a few' times over.}  Prior to such a time 
any density estimate would be largely meaningless, and talk of its uncertainty doubly so. 
Assuming that appropriate tests have been done to ensure that the above criterion is met, then the simple answer to `Question 2' is:\begin{quote}`Yes: a good estimator for $P$ is the transition matrix obtained by looking at the history $\vec t$ and finding within it the proportion of times with which it moved to bin $i$ given that it was already in bin $j$.'\end{quote}
Using the convention that $P$ is left-stochastic, we therefore take its estimator to have the value  
    $$\frac { \sum_k  
    \delta_{i, t_{k+1}}  \delta_{  j,{t_k}}
    }{s_j}$$ in its 
    $i^{\mathrm{th}}$ row and
    $j^{\mathrm{th}}$ column.
    Not only is this choice heuristically simple, these are also maximum likelihood estimators for the elements of $P$ (see discussions in \cite{norris_1997} and \cite{TREVEZAS20092242}). Is it `good enough'?  Again, we appeal to the the `equilibrium' assumption to argue that it is. While these will never represent the true elements of $P$,  we do not need them to do so. More quantitative arguments justifying this estimator for $P$ are found in \cite{TREVEZAS20092242}. 
    
What of Question 1?  To answer this we proceed in stages.
To begin with, we have proved (assuming as before that $P$ is left-stochastic) the 
intermediate result that:
\begin{align}
  \mathrm{Cov}[S_b,S_c]
    &= 
    N \pi_b(\delta _{bc} -\pi_c)
+
    \left\{
        \left( \frac{NQ}{\ident-Q}  - \frac{Q -Q^{N+1}}{(\ident-Q)^2}
    \right)_{cb}\pi_b 
    +[b\leftrightarrow c]\right\}
    \label{eq:allcovarapp}
\end{align}
within which a number of new symbols have been introduced.\footnote{We had to derive (\ref{eq:allcovarapp}) ourselves as we were unable to find it anywhere else in the literature, although its first few terms appear in many works (\cite{TREVEZAS20092242} is one) which calculate the limit, as $N\rightarrow\infty$, of $\mathrm{Cov}[S_b,S_c]/N$. We have not provided a proof of the result here, mostly for reasons of space, but also because the proof is not particularly illuminating and should be little more than an excercise for persons more familiar with Markov chains than we are.  Persons nontheless interested are welcome to contact the authors for further details.}
In particular: $Q$ is defined to be the matrix $Q=P-P^\infty$ in which $P^\infty$ is defined by $P^\infty=\lim_{n\rightarrow\infty} P^n$ which may itself be shown to be given by $$P^\infty=(\ident-P+U)^{-1} \cdot U$$ if $U$ is a $B\times B$ matrix consisting entirely of ones.  Furthermore, 
the $B$-vector $\vec \pi$ having components $\vec\pi=(\pi_1,\pi_2,\ldots,\pi_B)$ contains the limiting
probabilities for the chain to end up in any of its $B$ states.  $\vec\pi$ may be shown (see \cite{TREVEZAS20092242}) to be given by
$$\vec\pi = (\ident - P +U)^{-1} \cdot {\vec 1}$$ if $\vec 1$ is a $B$-vector consisting entirely of ones.
Finally we note that 
the repeated index $b$ on the right hand side of (\ref{eq:allcovarapp}) does not indicate the summation convention is being used.

Finally, we need to express the variances of the desired transformed quantities, $\mathrm{Var}[S''_b]$, in terms of the covariances of raw quantities, $\mathrm{Cov}[S_b,S_c]$, given in (\ref{eq:allcovarapp}).\footnote{This will be achieved in (\ref{eq:achievement}).} In principle this translation can be done for arbitrary functions $g_1, g_2, \ldots, g_B$, however in the interests of brevity our presentation here focuses on the {\it particular} transformation which is needed in this analysis to weight each of the raw
histogram counts before then normalising the histogram to unit area:
\begin{align}
    S''_b = \frac{S_b w_b}{\sum_{i=1}^B  S_i w_i } \label{relntodiff}
\end{align}
where the $\{w_1,w_2,\ldots,w_B\}$ are the set of fixed weights.
The trivial changes necessary to generalise to other transformations are left as an exercise for the reader! 
Taking differentials of (\ref{relntodiff}) and defining the fractional differentials $df_b$ and $df''_b$ by:
%
$$
d f_b =  \frac {d S_b} {S_b} 
\qquad\text{and}\qquad 
d f''_b =  \frac {d S''_b} {S''_b} 
$$
it may be proved that:
\begin{align}
    d f''_ b
    &= 
    \sum_{k=1}^B \left( 
     \delta_{bk}   
   - 
    S''_k  
    \right)  d f_k.
    \label{eq:thediffs}
\end{align}
To gain some understanding of (\ref{eq:thediffs}) we make some small notational changes.  We use the fact that $S''_b$ represents a probability estimate in bin $b$ to motivate re-labelling it as $p''_b$. And since complementary probabilities are often denoted with $q$s we define $q''_b=1-p''_b$.  With those changes, (\ref{eq:thediffs}) may be written out as:
\begin{align}
    d f''_ 1
    &= 
    + \underline{{\color{red} q''_1}} df_1 
    - {p''_2} df_2 
    - {p''_3} df_3 
    - \ldots - {p''_B} df_B 
    \\
    d f''_ 2
    &= 
    - {p''_1} df_1 
    + \underline{{\color{red} q''_2}} df_2 
    - {p''_3} df_3 - \ldots - {p''_B} df_B 
    \\
    d f''_ 3
    &= 
    - {p''_1} df_1 
    - {p''_2} df_2 
    + \underline{{\color{red} q''_3}} df_3
    - \ldots - {p''_B} df_B 
    \\
    & \phantom{=}\vdots \nonumber
    \\
    d f''_ B
    &= 
    - {p''_1} df_1 
    - {p''_2} df_2 
    - {p''_3} df_3 
    - \ldots 
    + \underline{{\color{red} q''_B}} df_B .
    \label{eq:thediffstwo}
\end{align}
Since the $p''$ and $q''$ quantities are all positive, in the above form one can see that the Markov chain correlations\footnote{Bootstrap based Markov chains are local and so will tend to positively correlate values of $df_b$ and $df_c$ when $b$ and $c$ are nearby, and anti-correlate them when far apart.} 
will tend to {\em reduce} the uncertainties in regions of high probability but have the opposite effect at large distances.  Putting everything together:
\begin{align}
    {\mathrm{ Var}}[
        S''_ b
    ] 
    &= 
    {\mathrm{ Var}}[d S''_b]
    = 
    {\mathrm{ Var}}[S''_b d f''_b]
     \approx
    s_b ^2
    {\mathrm{ Var}}[
        d f''_ b
    ] 
    \nonumber
    \\
    &= 
    s_b^2 {\mathrm {Var}}\left[
    \sum_{k=1}^B \left( 
     \delta_{bk}   
   - 
    S''_k  
    \right)  d f_k
    \right]\anote{by (\ref{eq:thediffs})}
\nonumber
\\
    &= 
    s_b^2
    \sum_{i , j} 
    {\mathrm {Cov}}\left[
        \left( \delta_{bi}   - S''_i  \right)  d f_i
    ,
        \left( \delta_{bj}   - S''_j  \right)  d f_j
    \right]
\nonumber
\\
    &\approx
    s_b^2
    \sum_{i, j} 
        \left( \delta_{bi}   - s''_i  \right)
        \left( \delta_{bj}   - s''_j  \right) 
    {\mathrm {Cov}}\left[
         d f_i
    ,
         d f_j
    \right]
\nonumber
\\
    &=
    s_b^2
    \sum_{i, j} 
        \left( \delta_{bi}   - s''_i  \right)
        \left( \delta_{bj}   - s''_j  \right) 
    {\mathrm {Cov}}\left[
        \frac{d S_i}{S_i}
    ,
         \frac{d S_i} {S_j}
    \right]
\nonumber
\\
    &\approx
    s_b^2
    \sum_{i, j} 
        \frac {\left( \delta_{bi}   - s''_i  \right) } {s_i}
        \frac {\left( \delta_{bj}   - s''_j  \right)  } {s_j}
    {\mathrm {Cov}}\left[
         d S_i
    ,
         d S_j
    \right]
\nonumber
\\
    &=
    s_b^2
    \sum_{i, j} 
        \frac {\left( \delta_{bi}   - s''_i  \right) } {s_i}
        \frac {\left( \delta_{bj}   - s''_j  \right)  } {s_j}
    {\mathrm {Cov}}\left[
         S_i
    ,
         S_j
    \right] \anote{which is computable using (\ref{eq:allcovarapp})}.\label{eq:achievement}
\end{align}
Each of the approximations above 
appeals to the `good enough' doctrine, previously discussed. Specifically, the above approximations are good if {\em fractional} uncertainties are `small', which is something that is already ensured by the previously used requirement that the Markov chain has begun to approach equilibrium, thereby placing uncertainties into a linear regime.

The above procedure is fast since the estimator for $P$ can be obtained during histogram filling at no extra cost, and the remainder of the steps only involve simple matrix operations on $B\times B$ matrices.

This concludes the description of how our uncertainties are estimated.  What remains to be understood is why this method of uncertainty estimation appears not to be in wider use in the high-energy physics community or, for that matter, elsewhere, despite the underlying maths being more than fifty years old. It would appear to have wide application in many places where, at present, we estimate uncertainties in Markov chain derived data by re-running the chains a few more times with different initial
conditions or random seeds.

\newcommand{\subf}[2]{%
  {\small\begin{tabular}[t]{@{}c@{}}
  #1\\#2
  \end{tabular}}%
}

\def\mainFigSanityCheckCurveOneNumEvents{200}
\def\mainFigSanityCheckCurveOneNumSamples{25000}
\def\mainFigSanityCheckCurveOneNumPreSamples{1000}
\def\mainFigSanityCheckCurveOneNumBins{47}
\def\mainFigSanityCheckCurveOneBinWidth{0.0004872340425531915}
\def\mainFigSanityCheckCurveOneTMin{-0.0027}
\def\mainFigSanityCheckCurveOneTMax{0.0202}
\def\mainFigSanityCheckCurveOneDistroName{distro\_uniform\_in\_cube}
\def\mainFigSanityCheckCurveOneSigmaLevel{(1,)}
\def\mainFigSanityCheckCurveOneDeltaInPhi{0.5}
\def\mainFigSanityCheckCurveTwoNumEvents{200}
\def\mainFigSanityCheckCurveTwoNumSamples{25000}
\def\mainFigSanityCheckCurveTwoNumPreSamples{1000}
\def\mainFigSanityCheckCurveTwoNumBins{47}
\def\mainFigSanityCheckCurveTwoBinWidth{0.0004872340425531915}
\def\mainFigSanityCheckCurveTwoTMin{-0.0027}
\def\mainFigSanityCheckCurveTwoTMax{0.0202}
\def\mainFigSanityCheckCurveTwoDistroName{distro\_uniform\_in\_cube}
\def\mainFigSanityCheckCurveTwoSigmaLevel{(1,)}
\def\mainFigSanityCheckCurveTwoDeltaInPhi{0.5}
\def\mainFigSanityCheckCurveThrNumEvents{200}
\def\mainFigSanityCheckCurveThrNumSamples{25000}
\def\mainFigSanityCheckCurveThrNumPreSamples{1000}
\def\mainFigSanityCheckCurveThrNumBins{47}
\def\mainFigSanityCheckCurveThrBinWidth{0.0004872340425531915}
\def\mainFigSanityCheckCurveThrTMin{-0.0027}
\def\mainFigSanityCheckCurveThrTMax{0.0202}
\def\mainFigSanityCheckCurveThrDistroName{distro\_uniform\_in\_cube}
\def\mainFigSanityCheckCurveThrSigmaLevel{(1,)}
\def\mainFigSanityCheckCurveThrDeltaInPhi{0.5}

\def\mainFigSanityCheckBorderlineCurveOneNumEvents{200}
\def\mainFigSanityCheckBorderlineCurveOneNumSamples{2500}
\def\mainFigSanityCheckBorderlineCurveOneNumPreSamples{1000}
\def\mainFigSanityCheckBorderlineCurveOneNumBins{30}
\def\mainFigSanityCheckBorderlineCurveOneBinWidth{0.0007633333333333333}
\def\mainFigSanityCheckBorderlineCurveOneTMin{-0.0027}
\def\mainFigSanityCheckBorderlineCurveOneTMax{0.0202}
\def\mainFigSanityCheckBorderlineCurveOneDistroName{distro\_uniform\_in\_cube}
\def\mainFigSanityCheckBorderlineCurveOneSigmaLevel{(1,)}
\def\mainFigSanityCheckBorderlineCurveOneDeltaInPhi{0.5}
\def\mainFigSanityCheckBorderlineCurveTwoNumEvents{200}
\def\mainFigSanityCheckBorderlineCurveTwoNumSamples{2500}
\def\mainFigSanityCheckBorderlineCurveTwoNumPreSamples{1000}
\def\mainFigSanityCheckBorderlineCurveTwoNumBins{30}
\def\mainFigSanityCheckBorderlineCurveTwoBinWidth{0.0007633333333333333}
\def\mainFigSanityCheckBorderlineCurveTwoTMin{-0.0027}
\def\mainFigSanityCheckBorderlineCurveTwoTMax{0.0202}
\def\mainFigSanityCheckBorderlineCurveTwoDistroName{distro\_uniform\_in\_cube}
\def\mainFigSanityCheckBorderlineCurveTwoSigmaLevel{(1,)}
\def\mainFigSanityCheckBorderlineCurveTwoDeltaInPhi{0.5}
\def\mainFigSanityCheckBorderlineCurveThrNumEvents{200}
\def\mainFigSanityCheckBorderlineCurveThrNumSamples{2500}
\def\mainFigSanityCheckBorderlineCurveThrNumPreSamples{1000}
\def\mainFigSanityCheckBorderlineCurveThrNumBins{30}
\def\mainFigSanityCheckBorderlineCurveThrBinWidth{0.0007633333333333333}
\def\mainFigSanityCheckBorderlineCurveThrTMin{-0.0027}
\def\mainFigSanityCheckBorderlineCurveThrTMax{0.0202}
\def\mainFigSanityCheckBorderlineCurveThrDistroName{distro\_uniform\_in\_cube}
\def\mainFigSanityCheckBorderlineCurveThrSigmaLevel{(1,)}
\def\mainFigSanityCheckBorderlineCurveThrDeltaInPhi{0.5}

\subsection*{Checks on uncertainties}

\begin{figure}
\centering
\includegraphics[width=0.9\textwidth]{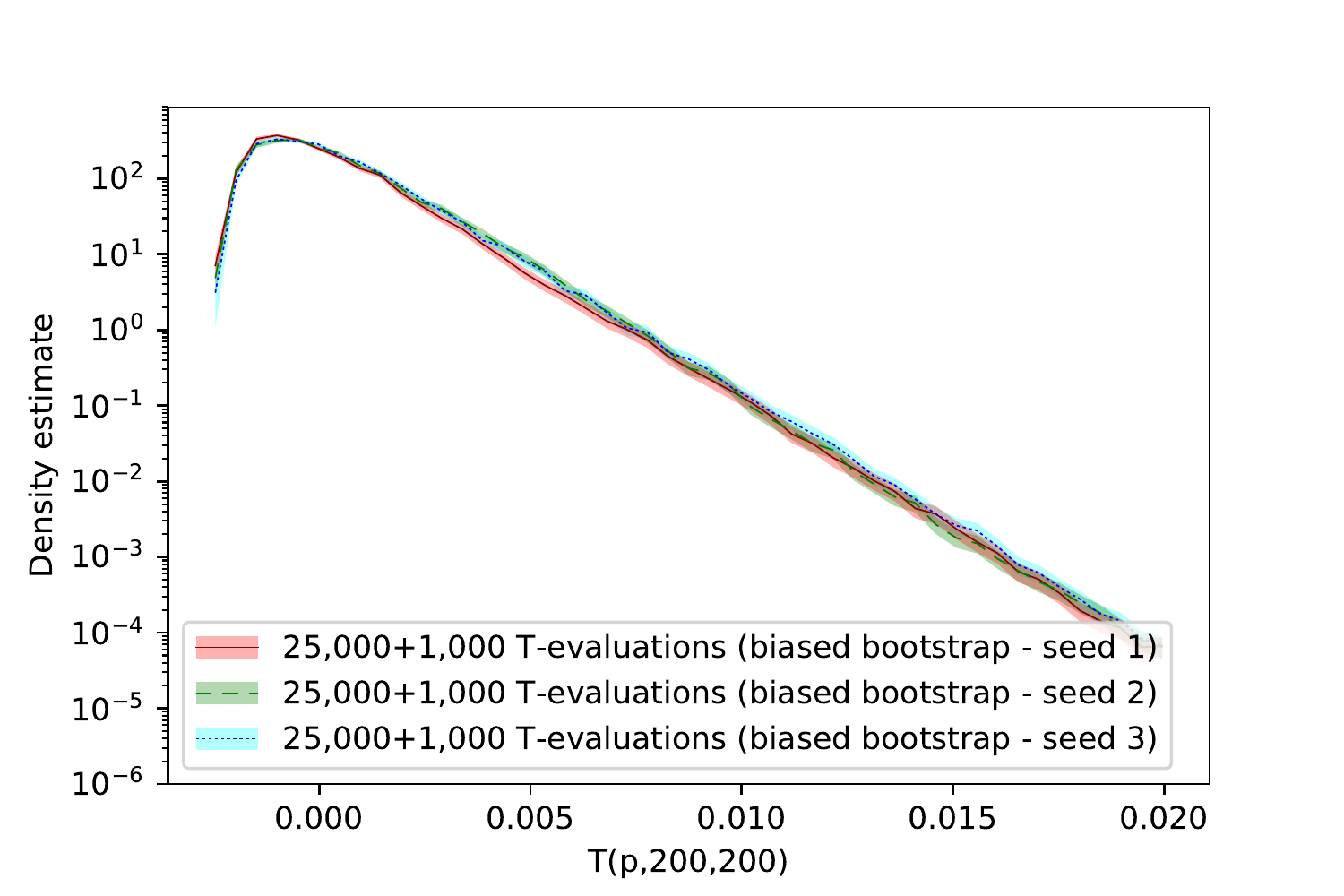}
    \caption{\label{fig:threehighstats} Comparison of three independent biased-bootstrap estimates of the density shown in Figure~\ref{fig:mainfiga}, together with their one-sigma uncertainty estimates.}
\end{figure}

Since our method of estimating uncertainties is both novel and non-trival, some readers may find it reassuring to see examples of it working as intended.  One way of doing this is to illustrate that independent density estimates from separate Markov chains agree within the uncertainty esitimates that this process attaches to each of them. We do so by presenting in Figure~\ref{fig:threehighstats} three independent density estimates which, apart from having different random number seeds, are otherwise identical to those shown in the biased-bootstrap estimate of Figure~\ref{fig:mainfiga}. In particular, all the estimates in Figure~\ref{fig:threehighstats} were generated from \num{\mainFigSanityCheckCurveOneNumSamples} biased bootstrap MCMC samples following \num{\mainFigSanityCheckCurveOneNumPreSamples} unbiased initialisation samples. Reassuringly, Figure~\ref{fig:threehighstats} does indeed show that these three estimates all have the degree of agreement with each other that would be expected given the sizes of their one-sigma uncertainty bands.

\begin{figure}
\centering
\begin{tabular}{ccc}
\hspace{-0.75cm}
\subf{\includegraphics[width=60mm]{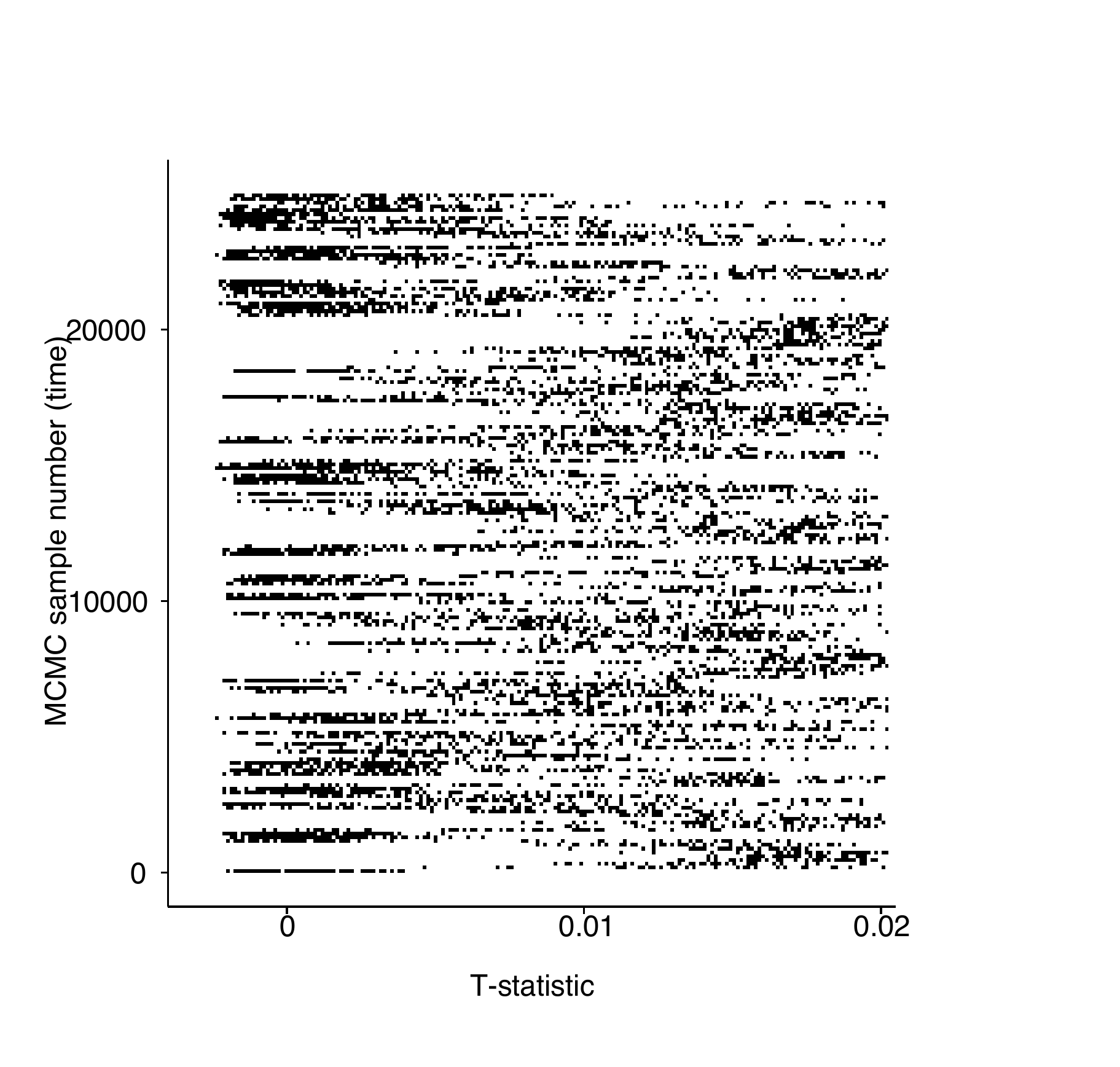}}
     {25,000 + 1,000 $T$-evaluations\\ (seed 1)}\hspace{-1.5cm}
&
\subf{\includegraphics[width=60mm]{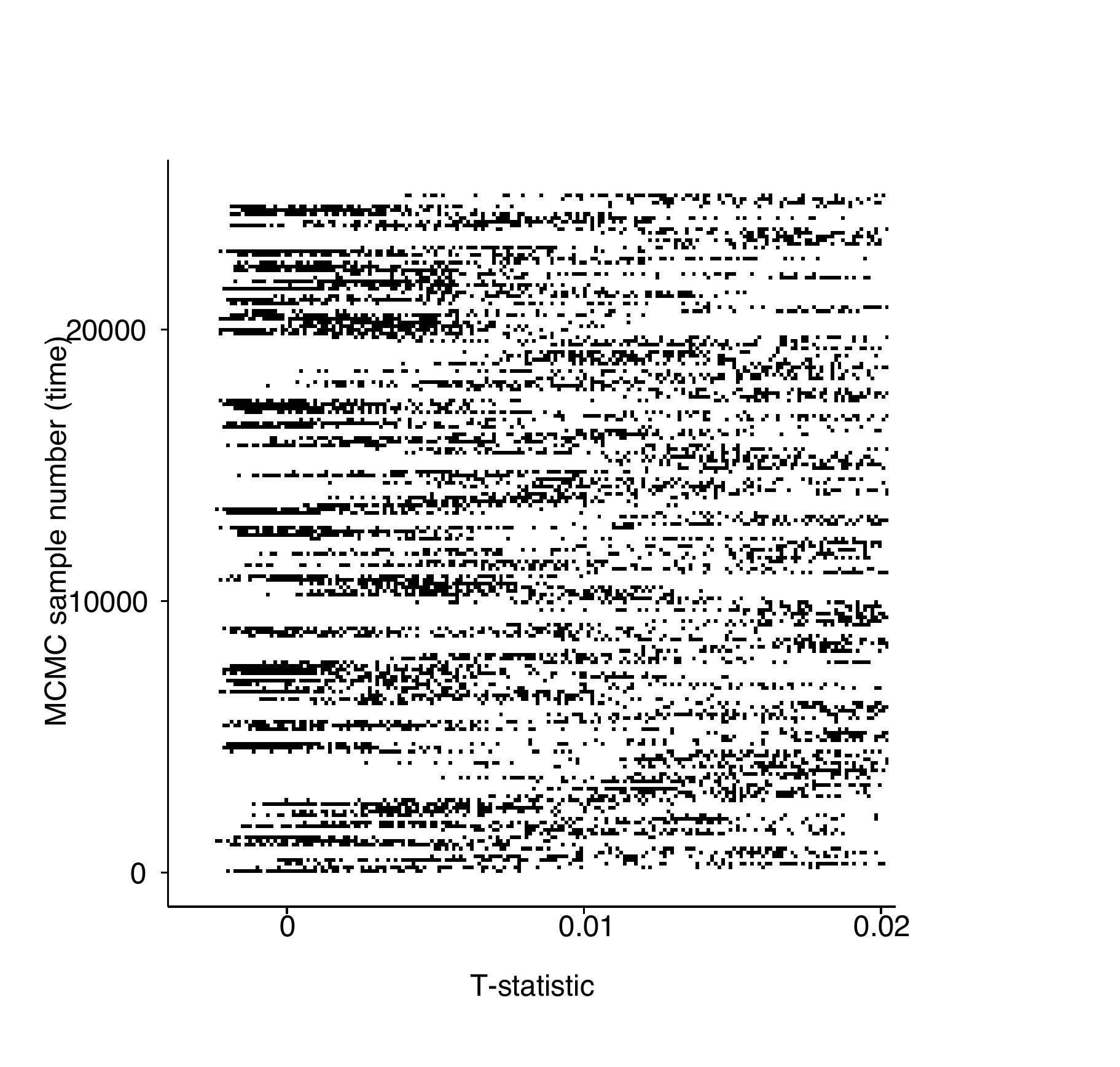}}
     {25,000 + 1,000 $T$-evaluations\\ (seed 2)}\hspace{-1.5cm}
&
\subf{\includegraphics[width=60mm]{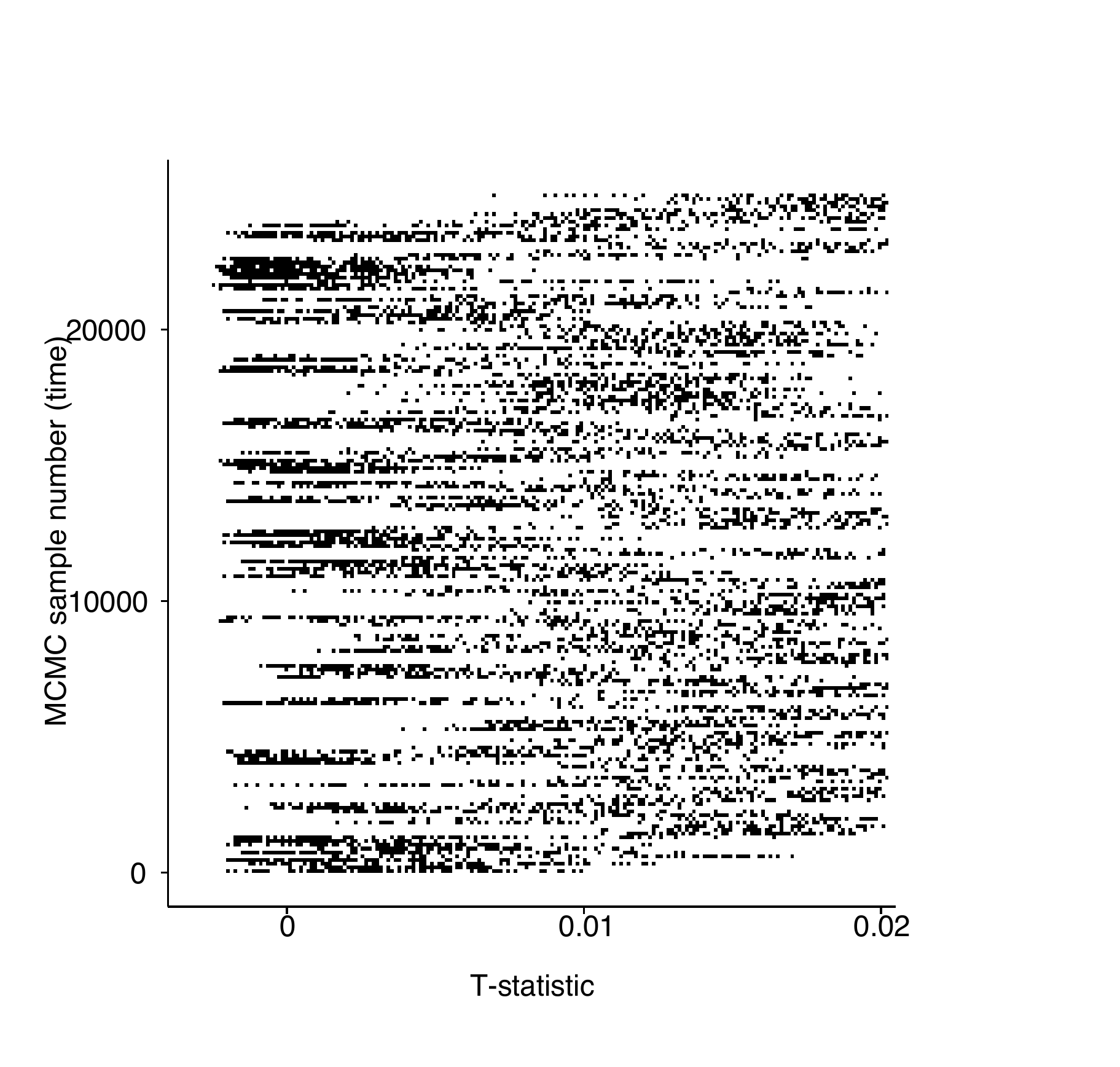}}
     {25,000 + 1,000 $T$-evaluations\\ (seed 3)}\hspace{-1.5cm}
\\
\hspace{-0.75cm}
\subf{\includegraphics[width=60mm]{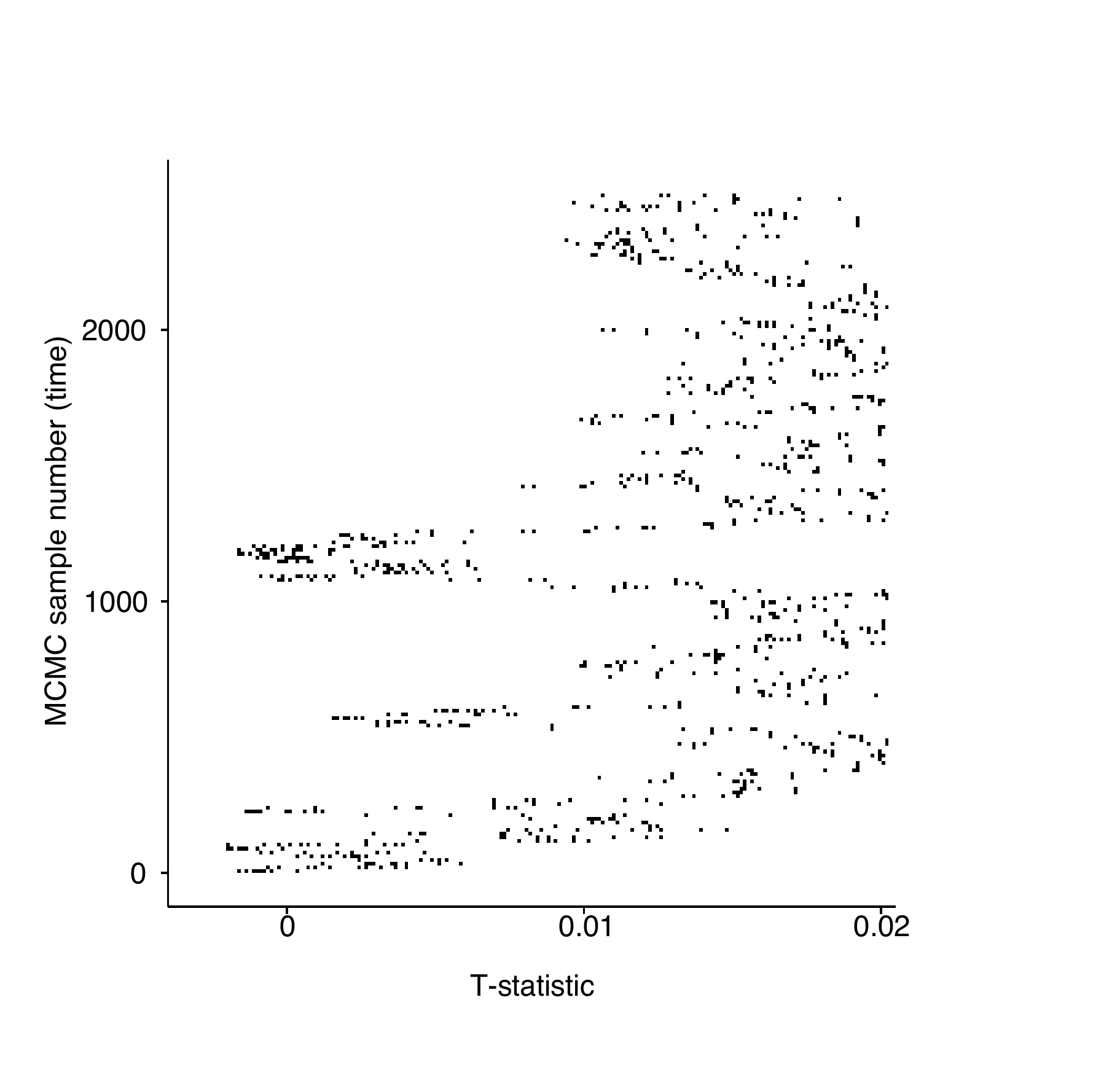}}
     {2,000 + 1,000 $T$-evaluations\\ (seed 1)}\hspace{-1.5cm}
&
\subf{\includegraphics[width=60mm]{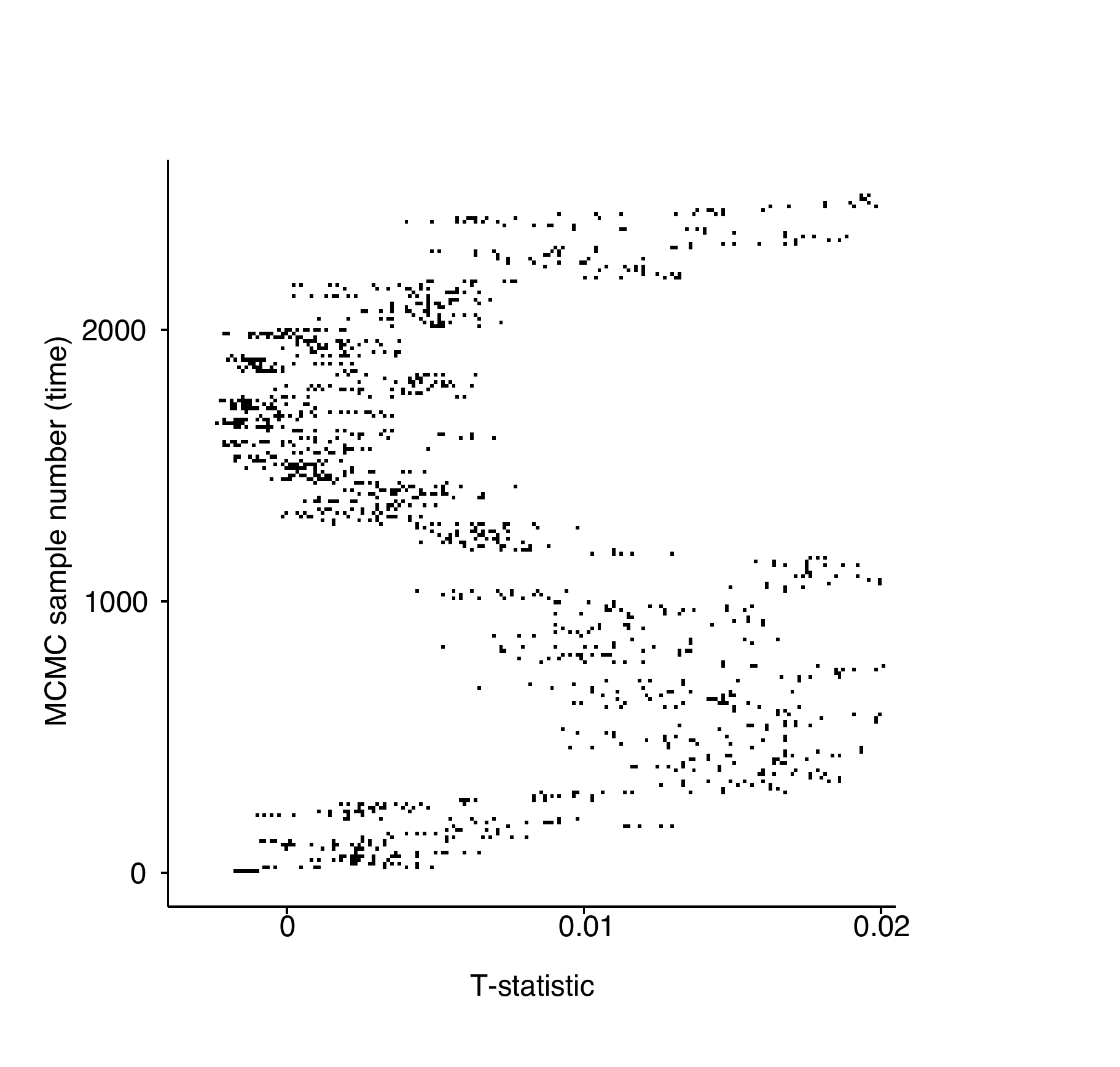}}
     {2,500 + 1,000 $T$-evaluations\\ (seed 2)}\hspace{-1.5cm}
&
\subf{\includegraphics[width=60mm]{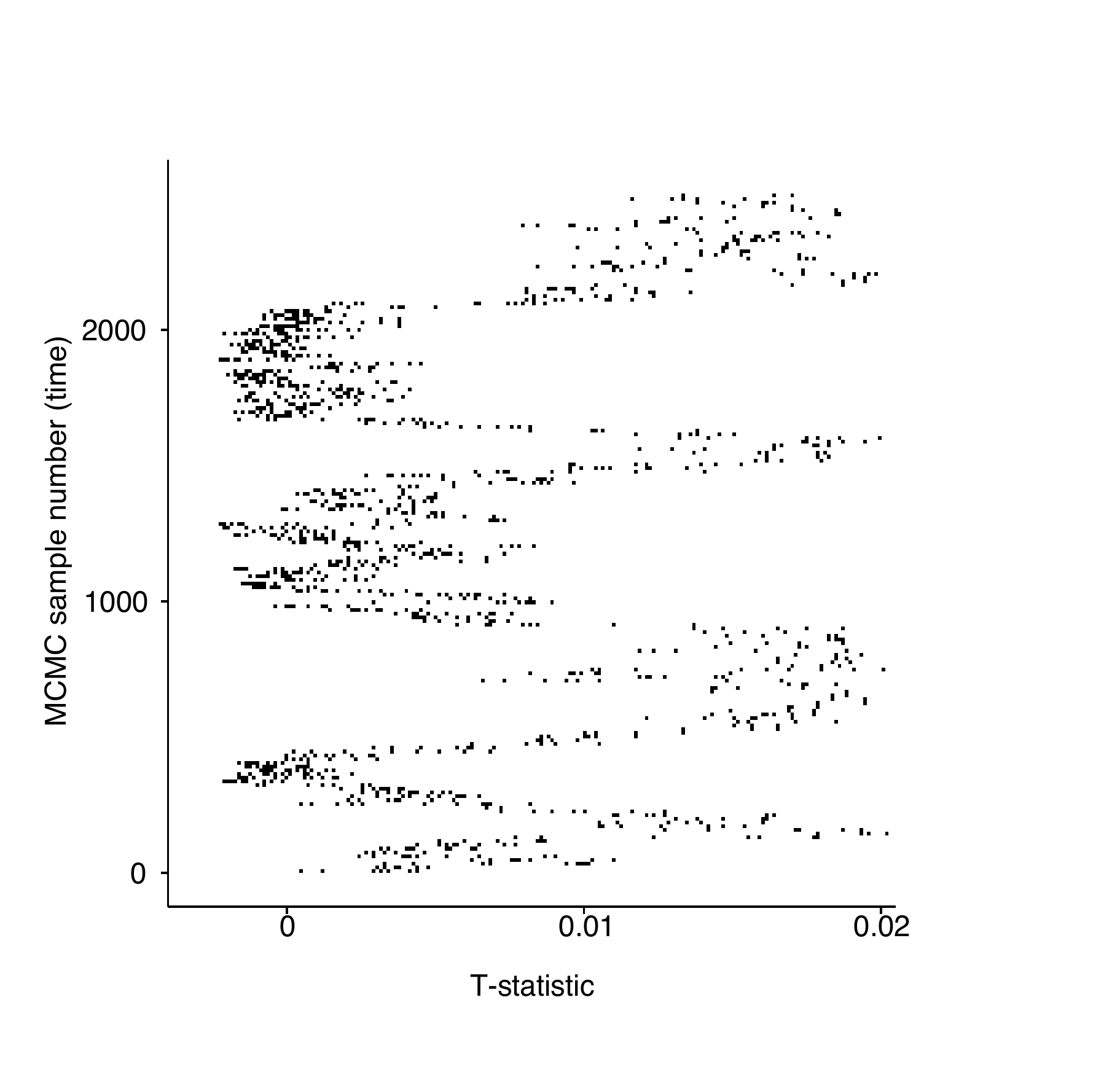}}
     {2,000 + 1,000 $T$-evaluations\\ (seed 3)}\hspace{-1.5cm}
\\
\end{tabular}
\caption{Histories of the $T$-values in three long and three short biased-bootstrap chains whose resulting density estimates are shown in Figures~\ref{fig:threehighstats} and \ref{fig:threeborderline}. \label{fig:histories}
}
\end{figure}

The top row of plots in Figure~\ref{fig:histories} show the history of the three chains which generated the densities just seen in Figure~\ref{fig:threehighstats}. In each case, time runs vertically.  The histories show: (a) expected auto-correlations within the chains, (b) that $T$-values were visited approximately uniformly, as desired, and (c) that the chains have made their way between the two ends of the $T$-spectrum $O(20)$ times.  The latter is strong evidence that the chain should have equilibriated and so the success of the uncertainty estimate ought not to be surprising.

\begin{figure}
\centering
\includegraphics[width=0.9\textwidth]{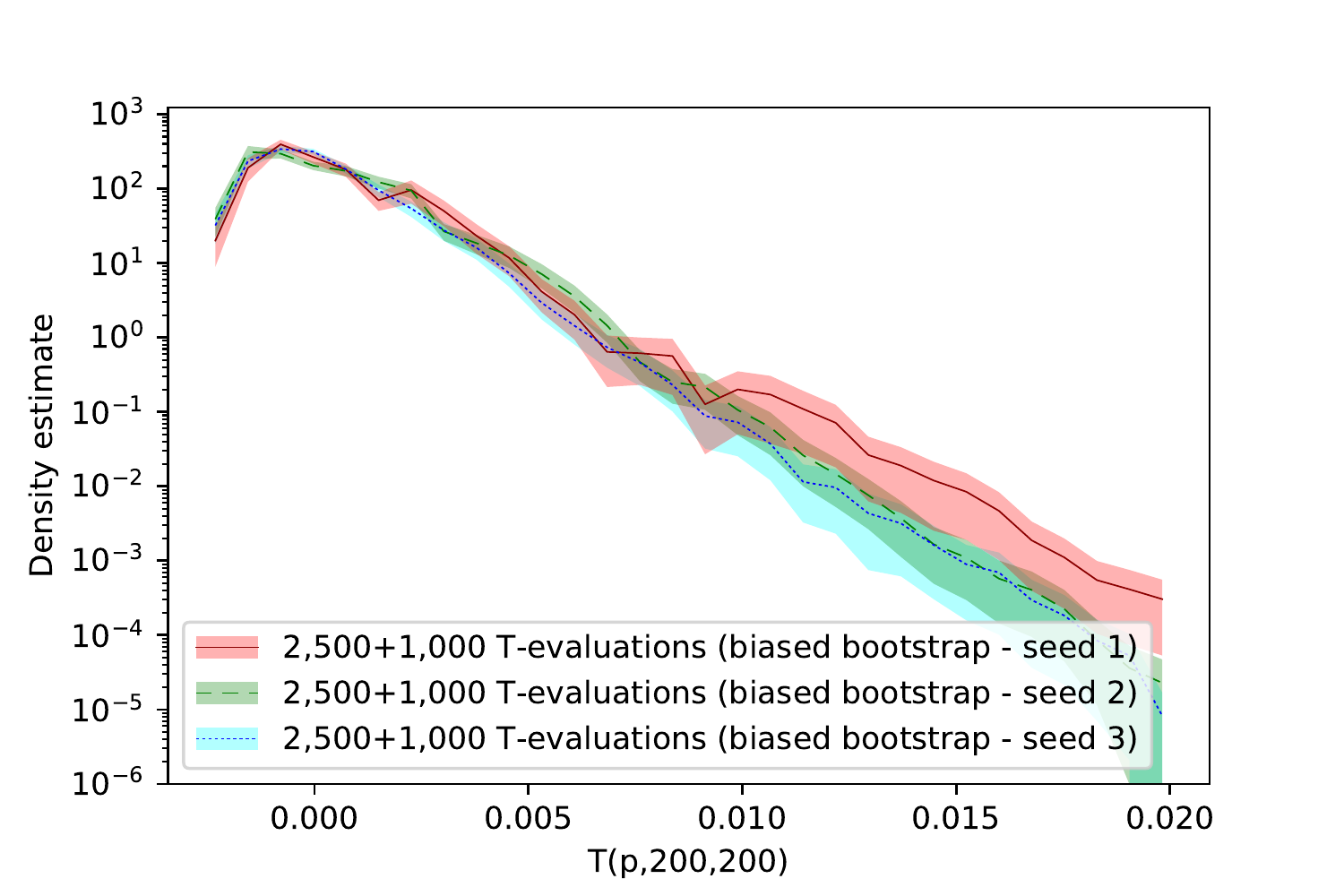}
    \caption{\label{fig:threeborderline} Comparison of three independent {\bf short-run} biased-bootstrap estimates of the density shown in Figure~\ref{fig:mainfiga}, together with their one-sigma uncertainty estimates. In contrast to earlier figures of a similar type, these density estimates here are calculated for $\mainFigSanityCheckBorderlineCurveOneNumBins$ bins on account of the lower statistics.
}
\end{figure}

What might be more interesting is to see the uncertainty estimate under tricker conditions.  The bottom row of plots in Figure~\ref{fig:histories} shows histories which are only 10\% of the length of those seen  previously (i.e.~2,500 MCMC biased-bootstrap samples instead of 25,000).  This time it is evident that these histories cannot be confidently said to have reached equilibrium. The first has hardly sampled any low $T$-values, while the second and third have gone from end to end only two or three times.  Nonetheless, despite these tough conditions, the one-sigma uncertainty estimates for these plots which are shown in Figure~\ref{fig:threeborderline} are once again remarkably well sized.  It is tempting to attribute this success to the unreasonable effectiveness of mathematics. 2,500 MCMC samples is sufficient to provide $O(100)$ samples in each of the 30 bins of Figure~\ref{fig:threeborderline}. The most likely transitions from each of those bins could therefore be relatively well estimated, and as a result of this the matrices $P$, $P^\infty$ and $Q$ can be much better known than one might at first imagine.  Since our method of uncertainty estimation is based on those matrices and analytically considers all possible MCMC chains consistent with them, not just the one (very short) chain that was actually realised in any of the three actualised histories, it is perhaps less surprising that the method succeeds so well. 

In conclusion: while our confidence in the method is primarily vested in the mathematics, we hope that these examples provide reassurance that our method of uncertainty estimation does indeed work well in the circumstances required -- namely any in which the esitimated $T$-distribution is remotely meaningful.

\end{appendices}
\bibliographystyle{alpha}
\bibliography{sample}

\end{document}